\documentstyle[12pt,twoside,fleqn,espcrc1,epsf]{article}
\def\la{\langle} 
\def\ra{\rangle} 
\def\be{\begin{eqnarray}} 
\def\ee{\end{eqnarray}}

\newcommand{\eq}{\begin{equation}} \newcommand{\eqx}{\end{equation}}

\newcommand{\eqn}{\begin{eqnarray}} \newcommand{\eqnx}{\end{eqnarray}}
\newcommand{\f}[2]{\frac{#1}{#2}}

\newcommand{\AmS}{{\protect\the\textfont2
  A\kern-.1667em\lower.5ex\hbox{M}\kern-.125emS}}

\title{A Four-Fermi Model in 0+1 Dimensions in Matter}

\author{Romuald A. Janik\address{Department of Physics, Jagellonian
        University, 
        30-059 Krakow, Poland.}, %
        Maciej A.  Nowak\addtocounter{address}{-1}\addressmark,
        G\'abor Papp\address{Institut f\"ur Theor. Physik, 
        Univ. Heidelberg, Philosophenweg 19,
         D-69120 Germany} %
    and
    Ismail Zahed\address{Department of Physics and Astronomy, 
        SUNY, Stony Brook, 
        New York 11794, USA.}}

\begin{document}
\maketitle

\begin{abstract}
The results of a number of constituent quark models in matter
may be understood in the mean-field approximation by using a 
simple four-fermi model in 0+1 dimensions. 
\end{abstract}

\section{Introduction}

The fate of the spontaneous breaking of chiral symmetry in matter
is still stirring a broad interest in light of present and future
heavy-ion collision experiments. Most of our understanding of the 
issue from first principles is limited to finite temperature, where
lattice simulations have been carried out~\cite{LATTICE}. At finite
chemical potential, present Monte-Carlo algorithms are upset by the
complex character of the measure~\cite{LOMBARDO}. 

A number of past and present analyses of the subject relies on QCD 
inspired models, such as the instanton model~\cite{INSTANTON,BOOK}, or
variants of the NJL model~\cite{ALL}. A common feature to
all these models, is the the emergence of chiral constituent quarks in the 
vacuum, following the spontaneous breaking of chiral symmetry. These models
do not confine.

The purpose of this talk is to show that the underlying mechanisms at
work in most of these models can be captured using a four-fermi model
in 0+1 dimensions~\cite{BOOK}. In section 1, we review the model
in the presence of chiral scalar and vector interactions. In section 2
we analyze its schematic thermodynamical content. In section 3, we 
construct the quark condensate and discuss its relation to the scalar
quark content of the constituent quark at low densities. In section 4,
we construct the resolvent and the distribution of eigenvalues for the
present model in the quenched approximation. In section 5, we derive the
spectral sum rules for the present model in the quenched and unquenched
approximation. Our conclusions are given in section 6.

\section{Model}

Consider a quark field $\psi_{a,f}$ 
where $a\!=\!1,2, ...,N$ are `color' indices, and $f\!=\!1,2, ...,N_f$
are flavor indices. For simplicity $N_f=1$ unless specified otherwise. 
By analogy with NJL models, the Lagrangian density is chosen to
be~\cite{JANIKNJL}
\be
  {\cal L}_{1}&=&
\label{model}
        \psi^\dagger\big(i\gamma_4\partial_4+im+i\mu\gamma_4\big)\psi \\
        &+&\f{g^2}2\Big((\psi^\dagger\psi)^2+
                (\psi^\dagger i\gamma_5\psi)^2\Big)+
        g^2_\omega (\psi^\dagger i\gamma_4\psi)^2 \nonumber
\ee
or equivalently
\be
  {\cal L}_{1}&=&
        R^\dagger i\big(\partial_4+
                (\mu\!-\!\omega_4)\big) L
        +\f 1{2g^2} PP^\dagger\!-\!\f 1{4g_\omega^2} \omega_4^2
                \nonumber \\
        &+& L^\dagger i\big(\partial_4 + 
                (\mu\!-\!\omega_4)\big)R 
\label{pboz} \\
        &+&R^\dagger i\big(P+m\big)R
                +L^\dagger i\big(P^\dagger+m\big)L \,. \nonumber \\
        \nonumber
\ee
The model will be discussed on a circle of radius $\beta=1/T$
in one dimension unless specified otherwise.
The gamma matrices are  chosen such that 
$\gamma_4\!=\!{\rm offdiag} (1,1)$ and $\gamma_5\!=\!{\rm diag} (1,-1)$,
in the chiral basis with $\psi=(R,L)$. The auxiliary fields are: 
$P\!=\!-2ig^2L^{\dagger}L$, $P^\dagger\!=\!-2ig^2 R^{\dagger} R$  and 
$\omega_4\!=\!-2 g_\omega^2\psi^\dagger i\gamma_4\psi$.

For $m=0$ (chiral limit), the model exhibits massive (constituent)
quark excitations at low densities, and massless (free) quark
excitations at high densities in the limit $N\rightarrow\infty$
(mean-field). In the same limit, $qq$, $qqq$, $\overline{q}q$, ... phases
and/or exitations are down by $1/N$. Although the model lacks confinement, it
bears much in common with more realistic chiral constituent quark models
such as the instanton or the NJL model~\cite{BOOK}. For an analysis of
the thermodynamics in a confining model we refer to~\cite{HANSSON}.

\section{Thermodynamics}

On a circle with boundary condition
$\psi (\tau+\beta)=-\psi(\tau)$, the operator $i\partial_4$ 
is invertible with a discrete spectrum $\omega_n=(2n+1)\pi T$.
In the mean-field approximation or large $N$, the pressure per particle
associated to (\ref{model}-\ref{pboz}) for $N_f=1$ is~\cite{BOOK,JANIKNJL}
\be
  \frac 1{\bf n}\, p=\omega -T\log\,(1-n)(1-\overline{n})
        -\Sigma PP^\dagger\!+\!\alpha\omega_4\omega_4^\dagger 
\label{pressureall}
\ee
with the usual occupancies $n,\bar{n}=1/(1+\exp((\omega\mp(\mu-\omega_4))/T))$.
Here $\omega=|P+m|$, $\Sigma=V_3/2g^2$ and $\alpha=V_3/g_{\omega}^2$,
following the rescaling $\psi\rightarrow\sqrt{V_3}\psi$. The thermodynamical
limit will be carried with $N\rightarrow\infty$ at fixed
${\bf n}=N/V_3$. (\ref{pressureall}) describes a quark with two energy
levels $\pm \omega$ at finite temperature and chemical potential. The
first term is the `zero point' motion, and the last two terms are exchange 
contributions~\cite{JANIKNJL}. The gap equations are
\be
  2\Sigma P = 1-n-\bar{n} \ ,\quad 
        \rho = {\bf n}\, (n-\bar{n})=2\alpha {\bf n}\,\,\omega_4 \,.
\label{gap1}
\ee
(\ref{gap1}) admit several solutions of which the one with 
maximum pressure will be selected. Generically there are two phases:
a broken phase with constituent quarks and a symmetric phase with free
quarks.
For $\alpha<\Sigma$ there is
a $\mu$ region with no real solutions to~(\ref{gap1}).
In the symmetric phase the pressure is $p\!={\bf n}\,(\mu\!-\!1/{4\alpha})$,
while in the broken  phase it is $p\!=\!{\bf n}/{4\Sigma}$, for $m=T=0$. The 
phase change occurs at $\mu_c=1/{4\Sigma}+ 1/{4\alpha}$. 

For $\alpha=\infty$ (vanishing vector coupling) the model is that used
originally by us~\cite{JANIKNJL}
and others~\cite{WETTIG}. In the broken phase $P_*=1/{2\Sigma}$ for 
$\rho=0$ with the pressure $p\!=\!{\bf n}\,(m\!+\!1/{4\Sigma})$. In the 
symmetric phase $P_*=0$ for $\rho= {\bf n}$ and  $p={\bf n}\mu$. The phase 
changes are mean-field driven and can be analyzed at the critical points 
using universality arguments~\cite{JANIKNJL}. 
For any $m$ there is a first order transition at $\mu_*\!=\!(m\!+\! 
1/{4\Sigma})$~\cite{BOOK,JANIKNJL}.

The occurrence of a first order transition for small values of the
temperature for light current masses is not generic to constituent
quark models. Indeed, in NJL models the first order transition in the
chiral limit is usually turned to a cross-over transition for light
quark masses~\cite{ALL}. The cross-over is robust against parameter
changes in the presence of a vector interaction. The first
order transition observed in these models at $T=0$ is much like the one 
studied in Walecka-type models in the form of a liquid-gas transition at 
low nucleonic matter density~\cite{KAPUSTA}. In the latter, the nucleon mass 
is the constituent mass and $m=0$.

\begin{figure}[tbp]
\centerline{\epsfxsize=85mm \epsfbox{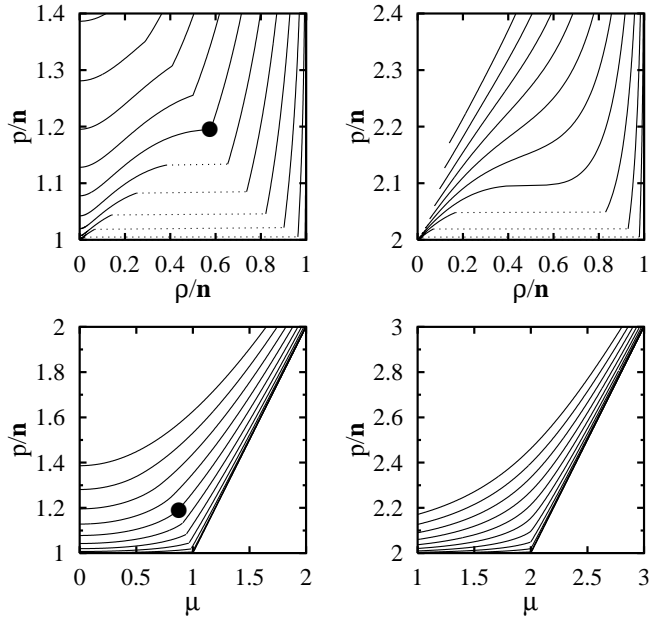}}
\caption{Isotherms at $m=0$ (left) and $m=1$ (right).}
\label{fig-diag}
\end{figure}
In Fig.~\ref{fig-diag} we show the isotherms in temperature
steps of $\Delta T=1/2$ for $m=0$ (left) and $m=1$ (right),
with $\Sigma=1$. At low temperature the transition is first order with 
a jump in the density (Maxwell construction) turning into a second order 
transition at high temperature. This is in agreement with our original 
arguments~\cite{BOOK,JANIKNJL}, and more recent investigations~\cite{NEW}. 
With increasing mass the `tricritical' point shifts down to lower temperatures,
and the second order transition is turned to a cross-over.
This confirms what is usually observed in NJL type models: a first order
transition in the massless case and a cross-over in the 
presence of light quarks~\cite{ALL}.

\section{Quark Condensate}

\begin{figure}[tbp]
\centerline{\epsfxsize=75mm \epsfbox{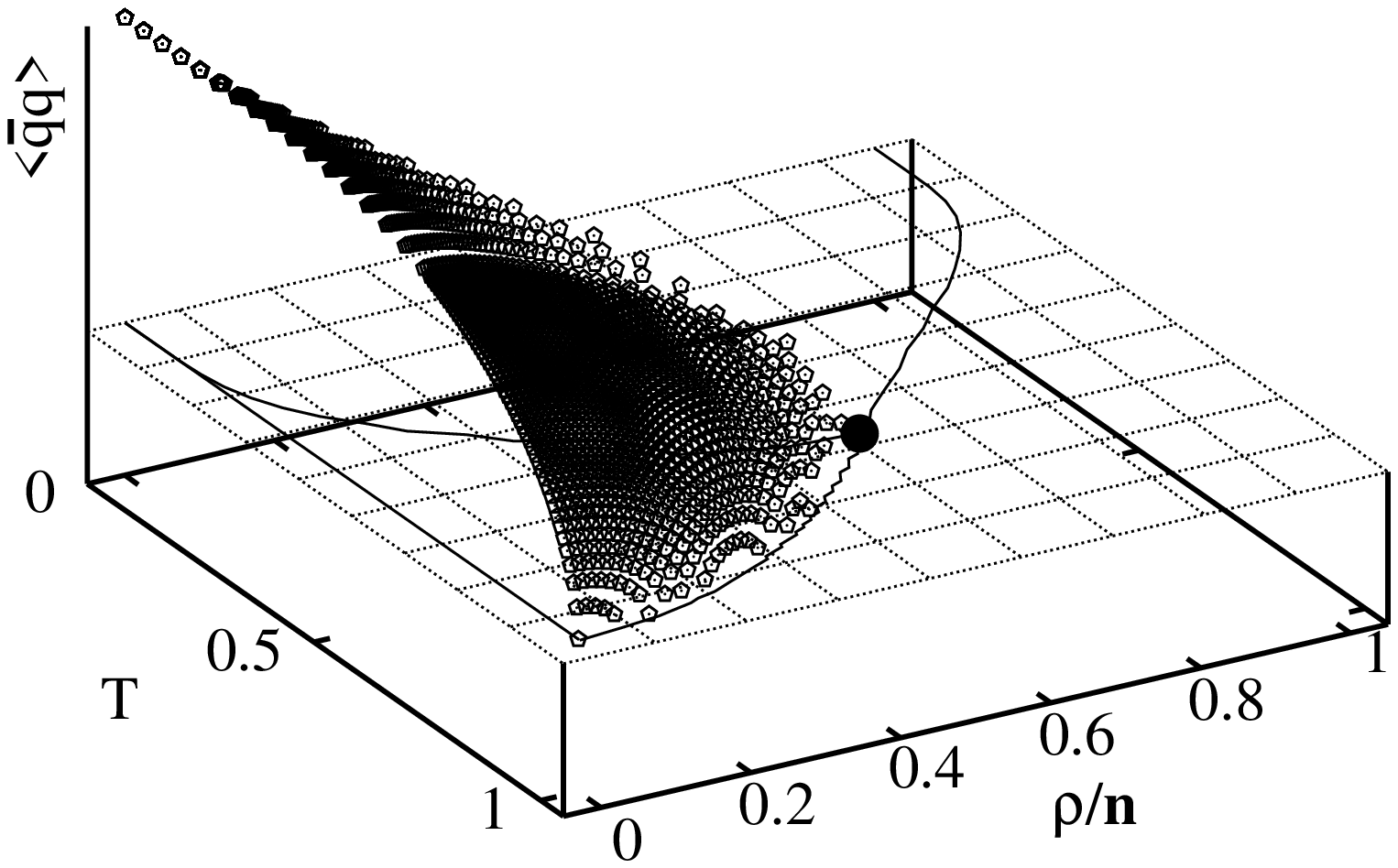}}
\caption{$|\la \overline{q}q\ra |$ in the $T\rho$ plane. See text.}
\label{fig-cond}
\end{figure}

The behavior of the quark condensate $|\la \overline{q} q\ra|=
2{\bf n}\,\Sigma m_Q$ with $m_Q=P_*$ and $m=0$
is shown in Fig.~\ref{fig-cond} as a function of temperature and density
in the chiral limit.
The solid line is the phase boundary crossing the tricritical point 
(big dot) beyond wich a mixed phase is developing. The hole in the middle
reflects on the mixed phase. The size of the hole shrinks with increasing mass
$m$, making it closer in shape to the result obtained in NJL type models 
in 4 dimensions~\cite{ALL}. Above the tricritical point $T_{3c}={m_Q}/3$ 
and $\rho_{3c}={\bf n}/{\sqrt{3}}$ or $\mu_{3c}={m_Q}\log(2\!+\!\sqrt{3})/3$, 
the transition is first order and disappears at $\mu_*\!=\!{m_Q}/2$. 
At $\mu=0$, the transition is second order (mean-field) and sets in at 
$T_*={m_Q}/2$. Amusingly, for a constituent mass $m_Q\sim 300$ MeV,
$\mu_*=T_*\sim 150$ MeV, and $\mu_{3c}\sim T_{3c}\sim 100$ MeV. In the
absence of matter, ${\bf n}=|\la \overline{q} q\ra|\sim (200\,{\rm MeV})^3$.
Hence $\rho_{3c}/\rho_0\sim 10/3$ where $\rho_0=0.17$ fm$^{-3}$ is nuclear
matter density. Most of these numbers are similar to the ones observed in
NJL models~\cite{ALL}.

The slope of the chiral condensate for small densities carries information
on the `pion-nucleon' $\sigma$ term. 
Here the role of the nucleons is played by constituent quarks.
For small
temperature the slope is $-1$ at the origin and reflects on the scalar 
charge of the constituent quark. This is to be compared to
$-2.5$ from the 2-flavour pion-nucleon sigma term with 
$\sigma_0\sim 40$ MeV and a current mass $m\sim 8$ MeV.

\section{Quark Spectrum}

For $\alpha=\infty$, the pseudoscalar 
four-Fermi interaction in (\ref{model}) 
causes the quarks to interact as if they were moving 
in a random Gaussian potential provided by the new auxiliary fields
${\cal A}_{ab} (\tau) \sim \psi_a \psi^{\dagger}_b (\tau)$ which is 
an $N\times N$ complex valued function of $\tau$\cite{BOOK,JANIKNJL}. 
For fixed ${\cal A}$, the quark spectrum follows from 
\be
\gamma_4 (i\partial_4 +{\cal A} +i\mu ) \psi^k
=\lambda_k [{\cal A}] \,\psi^k
\label{dirac}
\ee
with anti-periodic boundary conditions and
\be
2{\cal A} = {\bf A} \,(1+\gamma_5) + {\bf A}^{\dagger} \, (1-\gamma_5)\,.
\label{matrixA}
\ee
which is symmetric, complex and block-off-diagonal.
(\ref{dirac}) in the present model is the analogue of the
QCD eigenvalue equation in external gauge field. The Gaussian averaging
with moments
\be
\Big\la {\cal A}_{ab}{\cal A}^{\dagger}_{cd}\Big\ra_{\cal A}
 = \frac 1{2N\Sigma} 
\delta_{ac}\delta_{bd}
\label{weight}
\ee
restores the four-fermi interaction. The spectrum 
(in the $n=0$ sector for ${\cal A}$ symmetric)
can be readily probed through the distribution of eigenvalues
\be
\varrho (z, \overline{z}) =\f 1N \sum_k  \Big\la  \delta( z-\lambda_k
  [{\cal A}] )\Big\ra_{\cal A}
\label{density}
\ee
which is complex valued for finite $\mu$. 
Setting $T=0$ in~(\ref{pressureall}) the model dimensionally reduces to a
0+0 dimension one which is a matrix model. 

The quenched distribution (\ref{density}) can be easily constructed~\cite{STEPH,US}.
For small values of $\mu$ the density of 
eigenvalues concentrates near the imaginary axis, while for large values
it does not. The unquenched distribution of eigenvalues
is harder to generate (complex weight induced by quark feedback).
However, we expect qualitatively a similar behavior but with a finite
accumulation of eigenvalues on the real axis. The chiral
transition from finite density to zero set in at intermediate values of 
$\mu$ for which the density of eigenvalues along the real axis would 
vanish in the thermodynamical limit $N\rightarrow\infty$. To characterize the 
onset of the transition quantitatively we suggest to use the microscopic
sum rules as we now show.

\begin{figure}[tbp]
\centerline{\epsfxsize=85mm \epsfbox{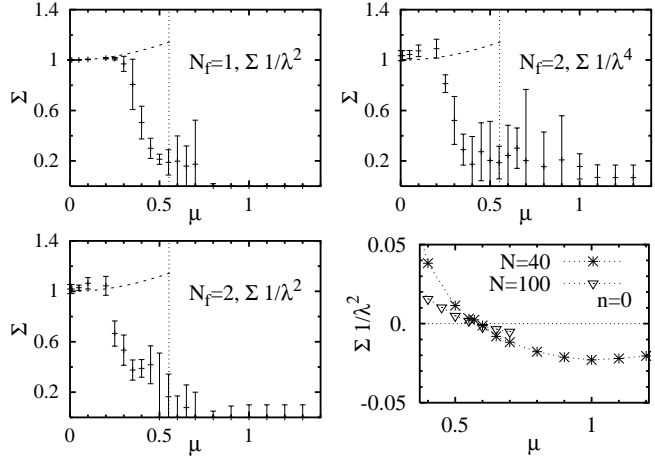}}
\caption{Spectral sum rules. See text.}
\label{fig-momp}
\end{figure}
\section{Spectral Sum Rules}

For fixed $N_f$, the spectral sum rule for our case reads
\be
\frac 1{N^2}
\Big\langle\Big\langle\sum_k'\frac 1{\lambda_k^2 [{\cal A}]}
\Big\rangle\Big\rangle_0 = \frac{\langle q^\dagger q\rangle^2}{2N_f}
        = \frac 1{2N_f} \left(\frac{\partial p}{\partial m}\right)^2
\label{smim4}
\ee
where the averaging is over the matrices ${\cal A}$
with the unquenched measure in the 0+0 model (no Matsubara modes).
The sum rule~(\ref{smim4}) allows for   
a determination of the chiral condensate (low density). Near $\mu\sim \mu_*$
the finite size corrections are important, upsetting the present construction.
In Fig.~\ref{fig-momp} the
chiral condensate extracted from~(\ref{smim4}) in the 0+0 model (averaged
over $5\times10^5$ matrices) is shown
for one (upper left) and two (lower left) flavours. In the upper right
corner we show the chiral condensate extracted from the next sum
rule. The dashed lines indicate the analytical result of the condensate.
The lower right corner shows the sum rule~(\ref{smim4}) for the quenched
($N_F=0$) case and {\em finite} $N$ (for $N\to\infty$ (\ref{smim4})
diverges). Due to the elongation of the support of eigenvalues for
large chemical
potential, the sum rule changes sign. The numerical analysis indicates
that this change occurs around the critical chemical potential
($\mu_*\approx 0.5277$~\cite{STEPH}). This suggests that the quenched physics
may still `remember' the unquenched one at the critical point, through the way 
the complex eigenvalues in ~(\ref{density}) get redistributed near zero.

%
%
%
%

\section{Conclusions}

We have constructed a chiral four fermi model
with the thermodynamical structure of a two-level quark system.
This model undergoes first and second order-type transitions, and 
can be used to illustrate how the quark spectrum may be probed in
matter. We used this model to show that the microscopic spectral sum rules
are sensitive to temperature and chemical potential changes including a 
phase transition in the unquenched approximation. The interplay between 
the thermodynamical 
limit, the chiral limit, the quenched-unquenched approximations  and the 
precision of numerical algorithms are readily addressed in this model.
We believe the present analysis to be useful for current lattice QCD 
simulations with matter.

\section{Acknowledgments}

GP and IZ  thank Gerry Brown  and Madappa Prakash for discussions.
This work was supported in part by the US DOE grant DE-FG-88ER40388, by the 
Polish Government Project (KBN) grants 2P03B04412, 2P03B00814 and
2P03B08614 and by the 
Hungarian grants FKFP-0126/1997 and OTKA-T022931.

\end{document}